\documentclass[12pt,reqno]{article}
\usepackage{amsfonts}
\usepackage{amsmath}
\usepackage{amsbsy}

\usepackage{amssymb,latexsym}
\usepackage{setspace}
 \numberwithin{equation}{section}

\begin{document}
 \allowdisplaybreaks[1]
\title{Non-linear Realisation of the $\mathcal{N}=2$, $D=6$ Supergravity}
\author{Nejat T. Y$\i$lmaz\\
Department of Mathematics
and Computer Science,\\
\c{C}ankaya University,\\
\"{O}\u{g}retmenler Cad. No:14,\quad  06530,\\
 Balgat, Ankara, Turkey.\\
          \texttt{ntyilmaz@cankaya.edu.tr}}
\maketitle
\begin{abstract}We have applied the method of dualisation
to construct the coset realisation of the bosonic sector of the
$\mathcal{N}=2$, $D=6$ supergravity which is coupled to a tensor
multiplet. The bosonic field equations are regained through the
Cartan-Maurer equation which the Cartan form satisfies. The
first-order formulation of the theory is also obtained as a
twisted self-duality condition within the non-linear coset
construction.
\end{abstract}

\textbf{Pacs numbers}: 04.65.+e, 12.60.Jv, 11.15.-q, 11.10.Lm.
\\


\section{Introduction}
The coset construction of the bosonic sectors of the maximal
supergravities obtained from the $D=11$ supergravity \cite{d=11}
by dimensional reduction has been given in \cite{julia2}. The
non-linear coset construction is based on the doubled formalism of
the theories in which a dual field is introduced for each bosonic
field of the theory. The results of \cite{julia2} have been
extended to the matter coupled supergravities in
\cite{nej4,nej5,nej6}. This has been possible since in
\cite{nej1,nej2} a general coset construction is performed for the
symmetric space sigma models of generic scalar coset manifolds.
Also in \cite{nej3} a general construction is presented for a
matter coupled scalar coset.

The coset constructions or the non-linear realisations of the
bosonic sectors of the supergravity theories are important for the
understanding of the global symmetries of these theories. The
global symmetry of the bosonic sector; in particular the scalar
sector can be extended over the other fields to be the rigid
global symmetry of the entire theory \cite{julia1,pope}. Besides a
restriction of the global symmetry group $G$ of a supergravity
theory to the integers is conjectured to be the U-duality symmetry
of the relative string theory whose low energy effective theory
becomes the supergravity theory at hand \cite{nej125,nej126}.

In \cite{nej7} the non-linear coset realisation of the pure
${\mathcal{N}}=4$, $D=5$ supergravity is given. In this work we
construct the coset realisation of the ${\mathcal{N}}=2$, $D=6$
supergravity which is coupled to a tensor multiplet
\cite{d=61,d=62}. The coupling of the tensor multiplet is
necessary to be able to write a Lorentz covariant lagrangian since
for the pure ${\mathcal{N}}=2$, $D=6$ graviton multiplet a
canonical Lorentz covariant lagrangian can not be constructed
owing to the existence of an anti-self-dual three-form field
strength \cite{d=61,d=62}. When one introduces a tensor multiplet
as it will be clear in section two one can lift the constraint of
(anti) self-duality by combining the fields of the graviton and
the tensor multiplets. In section two we will derive the field
equations and then we will give the locally integrated first-order
field equations. In section three following the introduction of
the coset map we will derive the algebra which parameterizes this
map by using the dualisation method of \cite{julia2}. We will
denote that the first-order field equations can be obtained from
the Cartan form which is induced by the coset map as a twisted
self-duality condition satisfied by it \cite{julia3}.
\section{The ${\mathcal{N}}=2$, $D=6$ Supergravity}
The field content of the pure ${\mathcal{N}}=2$, $D=6$
supergravity multiplet \cite{d=61,d=62} can be given as
\begin{equation}\label{21}
(e_{\mu}^{m},\:\psi_{\mu}^{i},\: A^{(-)}_{\mu\nu}),
\end{equation}
where $e_{\mu}^{m}$ is the vielbein, $A^{(-)}_{\mu\nu}$ is an
anti-self-dual two-form field and $\psi_{\mu}^{i}$ for $i=1,2$ are
the gravitini. Due to the anti-self-duality constraint on the
two-form gauge field $A$ there is no way of constructing a
canonical Lorentz covariant and unconstrained lagrangian for the
pure ${\mathcal{N}}=2$, $D=6$ supergravity
\cite{d=61,d=62,d=63,d=64,d=65}. The minimal coupling which
enables the construction of a Lorentz covariant lagrangian is the
coupling of a matter tensor multiplet whose field content is
\begin{equation}\label{22}
(\lambda_{i},\: A^{(+)}_{\mu\nu},\:\phi),
\end{equation}
where $A^{(+)}_{\mu\nu}$ is a self-dual two-form field, $\phi$ is
a scalar and $\lambda_{i}$ for $i=1,2$ are symplectic
Majorana-Weyl spinors. By field redefinitions one may combine the
anti-self-dual two-form $A^{(-)}_{\mu\nu}$ of the supergravity
multiplet and the self-dual two-form $A^{(+)}_{\mu\nu}$ of the
tensor multiplet into a single two-form field $A_{\mu\nu}$ which
is unconstrained and this enables the construction of the Lorentz
covariant and unconstrained lagrangian for the ${\mathcal{N}}=2$,
$D=6$ supergravity coupled to a tensor multiplet. Thus the field
content of the theory can be given as
\begin{equation}\label{23}
(e_{\mu}^{m},\:\psi_{\mu}^{i},\: A_{\mu\nu},\:\lambda_{i},\:
\phi).
\end{equation}
We assume the signature of the space-time metric as
\begin{equation}\label{24}
\eta_{AB}=\text{diag}(-,+,+,+,+,+).
\end{equation}
The bosonic lagrangian of the $\mathcal{N}=2$, $D=6$ supergravity
coupled to a tensor multiplet can be given as \cite{d=61,d=62}
\begin{equation}\label{25}
\mathcal{L}=-\frac{1}{2}\, R\ast 1 -\frac{1}{2}\, \ast d\phi\wedge
d\phi-\frac{1}{2}\, e^{2\phi}\,\ast F\wedge F,
\end{equation}
where $F=dA$. In the next section we will workout the coset
construction of the bosonic sector of the theory excluding the
gravity sector; for this reason we are interested in only the
equations of motion for the fields $A$ and $\phi$. If we vary the
lagrangian with respect to the fields $A$ and $\phi$ we find the
corresponding second-order field equations respectively as
\begin{subequations}\label{26}
\begin{gather}
d(e^{2\phi}\,\ast F)=0, \notag\\
\notag\\
d(\ast d\phi)=-e^{2\phi}\,\ast F\wedge
F.\tag{\ref{26}}\end{gather}
\end{subequations}
By using the fact that locally a closed differential form is an
exact one we can locally integrate the second-order field
equations to obtain the local first-order ones. Thus if we
introduce the three-form $\widetilde{A}$ and the four-form
$\widetilde{\phi}$, by eliminating an exterior derivative operator
on both sides of the field equations in \eqref{26} we can write
down the first-order equations as
\begin{subequations}\label{27}
\begin{gather}
e^{2\phi}\,\ast F=d\widetilde{A}, \notag\\
\notag\\
\ast d\phi=d\widetilde{\phi}+d\widetilde{A}\wedge
A.\tag{\ref{27}}\end{gather}
\end{subequations}
If one takes the exterior derivative of the equations in
\eqref{27} one would obtain the second-order field equations
\eqref{26} which are free of the lagrange multiplier fields
$\widetilde{A}$ and $\widetilde{\phi}$.
\section{The Doubled-formalism and the Coset Structure}
In this section we will construct a coset structure for the
bosonic sector of the $\mathcal{N}=2$, $D=6$ supergravity that is
coupled to a tensor multiplet such that the second-order field
equations of \eqref{26} can be realized by the Cartan form of the
coset map in the Cartan-Maurer equation
\cite{julia2,nej4,nej5,nej6,nej3}. The coset map will be
parameterized by a Lie superalgebra and our aim will be to derive
the structure of this algebra. We first assign the generators
\begin{equation}\label{31}
(\,Y,\: K\,),
\end{equation}
to the original bosonic fields $A$ and $\phi$ respectively. We
have already introduced the dual fields $\widetilde{A}$ and
$\widetilde{\phi}$ in \eqref{27} therefore we also define the dual
generators
\begin{equation}\label{32}
(\,\widetilde{Y},\,\widetilde{K}\,),
\end{equation}
which will couple to the dual fields in the construction of the
coset map. Since the Lie superalgebra of the generators of any
coset construction has a $\mathbb{Z}_{2}$ grading the generators
are chosen to be odd if the corresponding coupling field is an odd
degree differential form and otherwise even \cite{julia2}. For our
case all the generators defined in \eqref{31} and \eqref{32} are
even according to the above general scheme. In the construction of
the coset map we will make use of the differential graded algebra
structure of the differential forms and the field generators given
in \eqref{31} and \eqref{32}. The details of this algebra can be
referred in \cite{julia2,nej1,nej2,nej3,nej7}. Before giving the
definition of the full coset map which includes the dual fields
and the dual generators as well we define
\begin{equation}\label{33}
\nu = e^{\phi K}e^{AY}.
\end{equation}
If one assumes a matrix representation for the algebra generated
by the generators $K$ and $Y$, by using the matrix identities
\begin{equation}\label{34}
\begin{aligned}
de^{X}e^{-X}&=dX+\frac{1}{2!}[X,dX]+\frac{1}{3!}[X,[X,
dX]]+\cdot\cdot\cdot \:,\\
\\
e^{X}Ye^{-X}&=Y+[X,Y]+\frac{1}{2!}[X,[X,Y]]+\cdot\cdot\cdot \:,
\end{aligned}
\end{equation}
one can calculate the Cartan form
\begin{equation}\label{35}
\mathcal{G}=d\nu\nu^{-1},
\end{equation}
as
\begin{equation}\label{36}
\mathcal{G}=d\phi K+ e^{\phi} dA Y,
\end{equation}
where the only non-vanishing commutator of the algebra of $K$ and
$Y$ is defined as
\begin{equation}\label{37}
[K,Y]= Y.
\end{equation}
When we insert the Cartan form \eqref{36} in the Cartan-Maurer
equation
\begin{equation}\label{38}
d\mathcal{G}-\mathcal{G}\wedge\mathcal{G}=0,
\end{equation}
whose validity originates from the definition of $\nu$ in
\eqref{33} we see that we obtain the trivial Bianchi identities
\begin{subequations}\label{39}
\begin{gather}
d(e^{\phi}\, dA)= -e^{\phi}dA\wedge d\phi, \notag\\
\notag\\
d(d\phi)=0,\tag{\ref{39}}\end{gather}
\end{subequations}
for the field strengths $e^{\phi}dA$ and $d\phi$ of the potentials
$A$ and $\phi$ respectively. Thus we observe that in order to
obtain the realisation of the field equations \eqref{26} of the
theory one has to construct the coset map by using the dual fields
and the dual generators together with the original ones. Therefore
next we propose the map
\begin{equation}\label{310}
\nu^{\prime}=e^{\phi K}e^{AY}
e^{\widetilde{A}\widetilde{Y}}e^{\widetilde{\phi}\widetilde{K}}.
\end{equation}
The Cartan form $\mathcal{G}^{\prime}=d\nu^{\prime}\nu^{\prime-1}$
induced by this coset map will also satisfy the Cartan-Maurer
equation
\begin{equation}\label{311}
d\mathcal{G}^{\prime}-\mathcal{G}^{\prime}\wedge\mathcal{G}^{\prime}=0,
\end{equation}
canonically. Following the general method of the coset
construction \cite{julia2,nej4,nej5,nej6,nej1,nej2,nej3,nej7} we
will demand that when we calculate the Cartan form
$\mathcal{G}^{\prime}=d\nu^{\prime}\nu^{\prime-1}$ and insert it
in the Cartan-Maurer equation \eqref{311} we should reach the
second-order field equations \eqref{26} of the theory. One
immediately observes that the calculation of the Cartan form
$\mathcal{G}^{\prime}$ starting from the definition of the coset
map \eqref{310} needs the specification of the algebra structure
of the generators $Y,\: K,\: \widetilde{Y},\: \widetilde{K}$. As a
matter of fact this is the mechanism we need to derive the algebra
structure of the non-linear realisation. If one calculates the
Cartan form $\mathcal{G}^{\prime}$ in terms of the unknown
structure constants of the algebra of the generators by using the
identities \eqref{34} and then inserts it in \eqref{311}; by
comparing the result with the second-order field equations
\eqref{26} one can read the desired structure constants. We should
remark that to be able to use the identities \eqref{34} we assume
that we choose a matrix representation for the algebra generated
by the generators of \eqref{31} and \eqref{32}. Performing the
above mentioned calculation we find that the only non-vanishing
commutators of the algebra of the generators \eqref{31} and
\eqref{32} are
\begin{subequations}\label{312}
\begin{gather}
[K,Y]=Y\quad,\quad[K,\widetilde{Y}]=-\widetilde{Y},\notag\\
\notag\\
[Y,\widetilde{Y}]=\widetilde{K}.\tag{\ref{312}}
\end{gather}
\end{subequations}
Now by using the matrix identities \eqref{34}, also the
commutators of \eqref{312} we can calculate the Cartan form
$\mathcal{G}^{\prime}=d\nu^{\prime}\nu^{\prime-1}$ of the coset
map \eqref{310} explicitly as
\begin{equation}\label{313}
\begin{aligned}
\mathcal{G}^{\prime}&=d\nu^{\prime}\nu^{\prime-1}\\
\\
&=d\phi\,
K\: +\:e^{\phi}\,dA\, Y\: +\:e^{-\phi}\,d\widetilde{A}\: \,\widetilde{Y}\\
\\
&\quad +(\:d\widetilde{\phi}\: +\:A\,\wedge\, d\widetilde{A}\:
)\,\widetilde{K}.
\end{aligned}
\end{equation}
The coset construction of the supergravities also produces the
first-order formulation of these theories \cite{julia2,julia3}.
The locally integrated first-order field equations are encoded in
the doubled Cartan form $\mathcal{G}^{\prime}$ as a twisted
self-duality condition
\begin{equation}\label{314}
\ast\mathcal{G}^{\prime}=\mathcal{SG}^{\prime},
\end{equation}
which it satisfies \cite{nej4,nej5,nej6,nej7}. In \eqref{314}
$\mathcal{S}$ is a pseudo-involution of the algebra generated by
the generators in \eqref{31} and \eqref{32}. For our construction
we define its action on the generators as
\begin{subequations}\label{315}
\begin{gather}
\mathcal{S}Y=\widetilde{Y}\quad,\quad\mathcal{S}K=\widetilde{K},\notag\\
\notag\\
\mathcal{S}\widetilde{Y}=Y\quad,\quad\mathcal{S}\widetilde{K}=K.\tag{\ref{315}}
\end{gather}
\end{subequations}
The general construction of $\mathcal{S}$ for a generic coset
formulation can be referred in \cite{julia2,nej1,nej2}. Now, by
using the definition of $\mathcal{S}$ given in \eqref{315};
inserting \eqref{313} in \eqref{314} gives
\begin{subequations}\label{316}
\begin{gather}
e^{\phi}\,\ast dA=e^{-\phi}\, d\widetilde{A}, \notag\\
\notag\\
\ast d\phi=d\widetilde{\phi}+A\wedge
d\widetilde{A}.\tag{\ref{316}}\end{gather}
\end{subequations}
We see that these are the same equations with \eqref{27} of
section two which have been obtained from the second-order field
equations \eqref{26} by differential algebraic integration.
However, here they are obtained through the coset construction
which includes the definition of the coset map \eqref{310}, the
algebra structure of \eqref{312} whose generators parameterize the
coset map and the matrix representation chosen for this algebra.

As discussed in \cite{d=61,d=62} the ${\mathcal{N}}=2$, $D=6$
supergravity coupled to a tensor multiplet is the minimal
extension of the ${\mathcal{N}}=2$, $D=6$ supergravity multiplet
to write an invariant lagrangian. Accordingly we conclude that
from the coset construction point of view the algebra derived in
\eqref{312} is a minimal one thus it plays a special role in the
covariant Lagrangian formulation of the theory.
\section{Conclusion}
We have obtained the coset formulation of the bosonic sector of
the $\mathcal{N}=2$, $D=6$ supergravity which is coupled to a
tensor multiplet \cite{d=61,d=62}. We have derived the algebra
structure which is used to parameterize a coset map such that the
induced Cartan form realizes the second-order field equations in
the Cartan-Maurer equation. Thus the bosonic field equations of
the $\mathcal{N}=2$, $D=6$ supergravity coupled to a tensor
multiplet are obtained within the geometrical construction of the
non-linear sigma model. The first-order formulation of the theory
is also encoded in the coset construction. The locally integrated
first-order field equations can be found through a twisted
self-duality condition satisfied by the Cartan form
\cite{julia2,julia3}.

Our main objective in this work was to construct the algebra which
parameterizes the coset map and which generates the field
equations. The group theoretical structure of our coset
realisation can further be studied separately. In section three we
have stated that the algebra we have constructed can be considered
as a minimal one. This fact may be linked to the minimality of the
tensor multiplet coupling to write an invariant lagrangian. One
may inspect the role of the generators of the two-form field and
its dual in the algebra constructed in section three. After
studying the group theoretical construction of the coset one may
question what it means algebraically and geometrically to
introduce the other generators of the algebra to construct a model
which will enable a Lorentz covariant and an unconstrained
lagrangian.

The non-linear coset construction of this work can be extended to
include the gravity and the fermionic sectors. The dualisation of
the $\mathcal{N}=2$, $D=6$ supergravity which is coupled to other
multiplets can be studied to orient the position of the
$\mathcal{N}=2$, $D=6$ supergravity in the general dualisation
scheme of the supergravity theories and to improve our knowledge
of the global symmetries of these theories.

\end{document}